\documentclass[letterpaper,twocolumn,10pt,prb,aps,superscriptaddress,showpacs]{revtex4-1}

\usepackage{amsmath,amssymb,amsthm,amsfonts}
\usepackage[all]{xy}
\usepackage{graphicx}

\usepackage{color}
\definecolor{darkblue}{rgb}{0.,0.,0.4}
\definecolor{darkred}{rgb}{0.5,0.,0.}
\usepackage[pdftex,colorlinks=true,linkcolor=darkblue,citecolor=darkred,urlcolor=blue]{hyperref}

\newcommand{\ket}[1]{\left| #1 \right\rangle}
\newcommand{\bra}[1]{\left\langle #1 \right|}
\newcommand{\tr}{\mathop{\mathrm{Tr}}\nolimits}

\newcommand{\drawgenerator}[8]{%
\xymatrix@!0{%
& #8 \ar@{-}[ld]\ar@{.}[dd] \ar@{-}[rr] & & #7 \ar@{-}[ld]  \\%
#1 \ar@{-}[rr] \ar@{-}[dd] &  & #2 \ar@{-}[dd] &            \\%
& #6 \ar@{.}[ld] &  & #5 \ar@{-}[uu] \ar@{.}[ll]       \\%
#3 \ar@{-}[rr] &  & #4 \ar@{-}[ru]                       %
}%
}

\begin{document}

\title{Bifurcation in entanglement renormalization group flow of a gapped spin model}

\author{Jeongwan \surname{Haah}}
\affiliation{Department of Physics, Massachusetts Institute of Technology, Cambridge, MA 02139}
\affiliation{Institute for Quantum Information and Matter, California Institute of Technology, Pasadena, CA 91125}
\date{5 February 2014}

\begin{abstract}
We study entanglement renormalization group transformations for the ground states of a spin model, 
called cubic code model $H_A$ in three dimensions,
in order to understand long-range entanglement structure.
The cubic code model has degenerate and locally indistinguishable ground states under periodic boundary conditions.
In the entanglement renormalization,
one applies local unitary transformations on a state, called disentangling transformations,
after which some of the spins are completely disentangled from the rest and then discarded.
We find a disentangling unitary
to establish equivalence of the ground state of $H_A$ on a lattice of lattice spacing $a$
to the tensor product of ground spaces of two independent Hamiltonians $H_A$ and $H_B$
on lattices of lattice spacing $2a$.
We further find a disentangling unitary for the ground space of $H_B$ with the lattice spacing $a$
to show that it decomposes into two copies of itself on the lattice of the lattice spacing $2a$.
The disentangling transformations yield a tensor network description for the ground state of the cubic code model.
Using exact formulas for the degeneracy as a function of system size,
we show that the two Hamiltonians $H_A$ and $H_B$ represent distinct phases of matter.
\end{abstract}

\maketitle

\section{Introduction}

The renormalization group (RG) is a collection of transformations
that select out quantities relevant to long-distance physics.~\cite{Wilson1975RG}
It generally consists of averaging out short-distance fluctuations
and rescaling of the system in order to recover the original picture.
In practice, however, details of RG transformations are context-dependent.
When an action is given and the corresponding partition function is of interest,
the RG transformation concerns the effective parameters (e.g., coupling constants, temperature)
of the theory as a function of probing length/energy scale.
When a wave-function is of interest,
the RG transformation takes place in a parametrization space of the wave functions
such that the transformed wave-function recovers correlations at long distance.

This paper is on the wave function renormalization, focusing on long-range entanglement structure.
As the entanglement of many body system is not characterized by a single number,
our general goal is to compare states with well-known states or to classify them under a suitable RG scheme.~\cite%
{VerstraeteCiracLatorreEtAl2005Renormalization,Vidal2007ER,ChenGuWen2010transformation}
The entanglement between any adjacent pair of spins can be arbitrary
since it can be changed simply by applying a local unitary operator,
which will certainly not affect the long-range behavior in any possible way.
This means that we should allow local unitary transformations in our definition of equivalence of long-range entanglement,
and the block of spins on which the local unitary is acting should generally be regarded as a single degree of freedom;
the long-range entanglement will only depend on the entanglement among the coarse-grained blocks.
In the case where the state is represented by some fixed network of tensors,~\cite{VerstraeteCirac2004PEPS}
this observation has been used to choose the most relevant part of the
tensors~\cite{VerstraeteCiracLatorreEtAl2005Renormalization,ChenGuWen2010transformation}
and to speed up certain numerical calculations.~\cite{LevinNave2007Tensor}

Here, we study long-range entanglement of the ground states of a particular three-dimensional gapped spin model,
via local unitary transformations that simplify the entanglement pattern.
This model, called the cubic code model,~\cite{Haah2011Local} shares an important property with 
intrinsically topologically ordered systems,~\cite{Wen1991SpinLiquid}
namely the \emph{local indistinguishability}~\cite{BravyiHastings2011short} of ground states.
However, there are two crucial differences: One is that the degeneracy under periodic boundary conditions is a very sensitive function of the system size.
The other is that it only admits point-like excitations whose hopping amplitude is exactly zero in presence of any small perturbation.
Although the cubic code model as presented is exactly solvable,
it is important to ask for a corresponding continuum theory.
This is one of the main motivations of this work.

\begin{figure}
 \includegraphics[width=.45\textwidth]{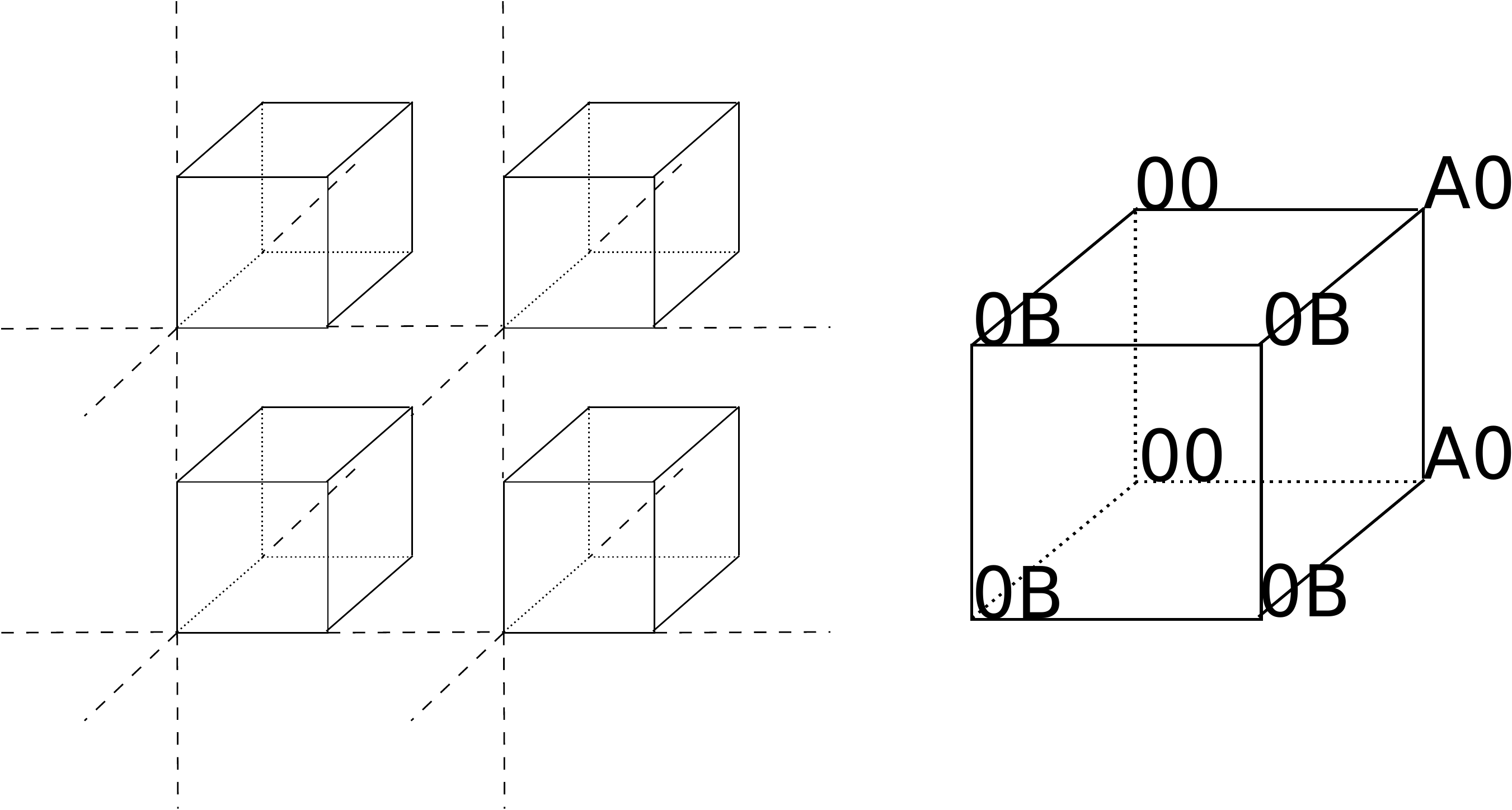}
 \caption{Simple cubic lattice of lattice spacing of $2a$ and the unit cell. There are 16 qubits labeled by $0$, $A$, or $B$ in the unit cell.
 Those that are labeled by $0$ are in the trivial product state. $\ket{\psi_A}$ and $\ket{\psi_B}$ in Eq.~\eqref{eq:state-A-to-AB}
 are states of the system of the qubits labeled by $A$ and $B$, respectively.}
 \label{fig:unit-cell}
\end{figure}

Our result is as follows. Let $H_A(a)$ be the Hamiltonian of the cubic code model. (See Eq.~\eqref{eq:H-model-A}.)
$H_A(a)$ lives on a simple cubic lattice with two qubits per site where the lattice spacing is $a$.
(We will mostly use the term `qubit' in place of `spin-$1/2$' from now on,
since only the fact that each local degree of freedom is two-dimensional is important.)
Let $H_B(a)$ denote another gapped spin Hamiltonian on a three-dimensional simple cubic lattice with four qubits per site
where the lattice spacing is $a$. $H_B(a)$ will be given explicitly later in Eq.~\eqref{eq:H-model-B}.
We find a constant number of layers of local unitary transformations (finite-depth quantum circuit) $U$ 
such that for any ground state $\ket{\psi_A(a)}$ of $H_A(a)$, we have
\begin{equation}
U \ket{\psi_A(a)} = \sum_i c_i \ket{\psi_A^i (2a)}\otimes \ket{\psi_B^i(2a)} \otimes \ket{ \uparrow \cdots \uparrow }
\label{eq:state-A-to-AB}
\end{equation}
where $c_i$ are complex numbers that depend on $\ket{\psi_A(a)}$, and $\ket{\psi_A(2a)^i},~\ket{\psi_B(2a)^i}$ are
ground states of $H_A(2a),~H_B(2a)$, respectively.
Note that on the right-hand side the wave function lives on the coarser lattice with lattice spacing $2a$.
The coarser lattice is depicted in Fig.~\ref{fig:unit-cell}. 
The unit cell of the coarser lattice has 16 qubits per Bravais lattice point.
10 qubits in each unit cell are in the trivial state, disentangled from the rest.
The Hamiltonian $H_A$ and $H_B$ live on the disjoint systems of qubits designated by $A$ and $B$ in Fig.~\ref{fig:unit-cell},
respectively.

Furthermore, we find another finite-depth quantum circuit $V$ such that for any ground state $\psi_B(a)$ of $H_B(a)$,
\begin{equation}
 V \ket{\psi_B(a)} = \sum_i c'_i \ket{\psi_B^i (2a)} \otimes \ket{\psi_B^i(2a)} \otimes \ket{ \uparrow \cdots \uparrow }
 \label{eq:state-B-to-BB}
\end{equation}
for some numbers $c'_i$. Again, on the right-hand side the wave functions live on the coarser lattice.
The qubits in the trivial state in Eq.~\eqref{eq:state-B-to-BB} are uniformly distributed throughout the lattice, similar to Fig.~\ref{fig:unit-cell}.
The first and second $\ket{\psi_B}$ in Eq.~\eqref{eq:state-B-to-BB} are states of disjoint systems of qubits,
similar to $A$ and $B$ of Fig.~\ref{fig:unit-cell}.

The result can be written suggestively as
\begin{equation}
 \mathcal R ( H_A ) = H_A \oplus H_B, \quad \mathcal R ( H_B ) = H_B \oplus H_B
 \label{eq:rg-suggestive}
\end{equation}
where $\mathcal R$ denotes the disentangling transformation followed by the scaling transformation by a factor of 2.
This is rather unexpected and should be contrasted 
with the previous results.~\cite{AguadoVidal2007Entanglement,ChenGuWen2010transformation,%
Aguado2011,BuerschaperMombelliChristandlEtAl2013,SchuchCiracPerez-Garcia2010G-injective}
It has been known that Levin-Wen string-net model~\cite{LevinWen2005String-net} and Kitaev quantum double model~\cite{Kitaev2003Fault-tolerant}
are entanglement RG fixed points.
Those results would have been summarized as $\mathcal R (H) = H$.
The ground-state subspace is retained at the coarse-grained lattice.
There was no splitting.
We will comment further on it later.

The present paper is organized as follows.
We begin by defining the model and reviewing its properties in Sec.~\ref{sec:model}.
We give details on the entanglement RG in Sec.~\ref{sec:erg}.
The actual unitary operators appearing in Eqs.~\eqref{eq:state-A-to-AB},\eqref{eq:state-B-to-BB}
will not be displayed in the text, but in a Mathematica script in Supplementary Material.~\cite{SM}
Next, we argue in Sec.~\ref{sec:diff-phase} that the newly found Hamiltonian $H_B$ represents a different phase of matter,
based on the degeneracy formulas of the models on periodic lattices.
In Sec.~\ref{sec:bMERA}, we point out the relevance of so-called branching MERA~\cite{EvenblyVidal2012RG}
description for the ground states of the cubic code model.
In Sec.~\ref{sec:method}, we describe a special representation of the models,
exploiting the translation symmetry and properties of Pauli matrices.
The special representation simplifies the calculation 
of the unitaries of Eqs.~\eqref{eq:state-A-to-AB},\eqref{eq:state-B-to-BB} significantly.
Sec.~\ref{sec:alg-check} builds on the preceeding section,
giving an algebro-geometric criterion and some intuition behind the entanglement RG calculations.
We conclude with a short discussion in Sec.~\ref{sec:discussion}.
Appendix~\ref{sec:entropy-bound} contains a direct bound~\cite{EvenblyVidal2013bMERAentropy}
on the entanglement entropy of a branching MERA state for a box region.

\section{Model}
\label{sec:model}

The spin model primarily considered in this paper
is described by an unfrustrated translation-invariant Hamiltonian
on the simple cubic lattice $\Lambda = \mathbb Z ^3$ with two qubits per lattice site.~\cite{Haah2011Local}
\begin{align}
 H_A = - J \sum_{i \in \Lambda} \left( G^x_i + G^z_i \right) \label{eq:H-model-A}
\end{align}
where $J > 0$ and
\begin{align}
 G^x_i &= \sigma^x_{i,1} \sigma^x_{i,2} \sigma^x_{i+\hat x,1} \sigma^x_{i+\hat y,1} \sigma^x_{i+\hat z,1} 
         \sigma^x_{i+\hat y+\hat z,2} \sigma^x_{i+\hat z+\hat x,2} \sigma^x_{i+\hat x+\hat y,2} \\
 G^z_i &= \sigma^z_{i,1} \sigma^z_{i,2} \sigma^z_{i-\hat x,2} \sigma^z_{i-\hat y,2} \sigma^z_{i-\hat z,2} 
         \sigma^z_{i-\hat y-\hat z,1} \sigma^z_{i-\hat z-\hat x,1} \sigma^z_{i-\hat x-\hat y,1}
\end{align}
are eight-qubit interaction terms consisted of Pauli matrices.
The index $i$ runs over all elementary cubes.
The terms $G^x_i$ and $G^z_i$ are visually depicted as
\[
\drawgenerator{IX}{II}{XI}{IX}{XI}{XX}{IX}{XI}
\drawgenerator{IZ}{ZZ}{ZI}{IZ}{ZI}{II}{IZ}{ZI} 
\xymatrix@!0{%
 & & \\
 & \circ \ar[u]^{\hat z} \ar[ld]^{\hat x} \ar[r]^{\hat y} & \\
 &                      &
}
\]
For the arrangement of the Pauli matrices on the vertices of the unit cube,
this is called {\em cubic code model}.
(It is a quantum error correcting code, but we will not discuss the theory of quantum error correction.)
One can easily verify that each term $G^x_i$ or $G^z_i$ commutes with any other term 
$G^x_j$ or $G^z_j$ in the Hamiltonian $H_A$.
A ground state $\ket \psi$ of $H_A$ can be written as
\begin{align}
 \ket \psi = \sum_{G \in \mathcal G} G \ket{\uparrow \uparrow \cdots \uparrow}  \label{eq:gs}
\end{align}
where $\mathcal G$ is the abelian group generated multiplicatively by terms $G^x_i$'s and $G^z_i$'s.
Since $ \ket{\uparrow \uparrow \cdots \uparrow}$ is an eigenstate of $G^z_i$ for any $i$ with eigenvalue $+1$,
the group $\mathcal G$ can be replaced by a smaller group consisting all products of $G^x_i$'s.
The ground state is degenerate (ground space). This will not concern us.

The energy spectrum can be understood by commutation relations among Pauli matrices,
since the Hamiltonian Eq.~\eqref{eq:H-model-A} is a sum of commuting tensor products of Pauli matrices.
Let us call a tensor product of Pauli matrices $\sigma^x, \sigma^y, \sigma^z$
 a {\em Pauli operator}.
If $\ket \psi$ is a ground state and $P$ is any Pauli operator,
then $P \ket \psi$ is also an energy eigenstate.
In fact, it is a common eigenstate for $G^x_i$ and $G^z_i$.
This is because any term $G^x_i$ or $G^z_i$ in the Hamiltonian, being a Pauli operator,
either commutes or anticommutes with $P$ ($P G^{x,z} = \pm G^{x,z} P$) 
and the ground state $\ket \psi$ is stabilized by any $G^x$ and $G^z$ ($G^{x,z} \ket \psi = \ket \psi$).

To understand the (excited) state $P \ket \psi$ better, imagine that
we measure all $G^x$ and $G^z$ simultaneously.
This is possible since they are pairwise commuting.
The measurement outcomes on $P \ket \psi$ are definite and take values $\pm 1$.
Let us say that there is a $X$-type \emph{defect} at $i$ if the expectation value of $G^x_i$ is $-1$.
Likewise, we define $Z$-type defects. Each defect has energy $2J$, and a state with no defect is a ground state.
A configuration of the defects characterizes an excited state effectively,
but not uniquely due to the ground state degeneracy;
for orthogonal ground states $\ket \psi$ and $\ket \phi$,
orthogonal states $P \ket \psi$ and $P \ket \phi$ give the same configuration of defects.
Note that the whole Hilbert space
is spanned by states of form $P \ket \psi$ for some Pauli operator $P$ and
some ground state $\ket \psi$.

An exotic property of the cubic code model is that
the excitations are \emph{pointlike} but \emph{immobile}.
They are pointlike because a single isolated defect is a valid configuration,
but are immobile because they are not allowed to hop to other position by a local operator.
Here, the locality is important.
There indeed exists a non-local operator that annihilates a defect
and create another at a different place.
The statement remains true even if we loosen our restriction that there be exactly one defect at $p$.
In a general case, one should distinguish a cluster of defects that is locally created,
in which case we call the cluster \emph{neutral},
from a cluster that is not locally created, in which case we call the cluster \emph{charged}.
(Since the charged cluster has nothing to do with any symmetry,
it is termed ``topologically charged.'')
The immobility asserts that any charged cluster
cannot be transported by any operator of finite support.

Rigorously, the immobility is stated as follows.
Suppose $\ket \psi$ is a state with a single defect, or more generally any charged cluster of defects,
contained in a box of linear dimension $w$.
Let $\mathbb{T}$ denote a translation operator by one unit length in the lattice along arbitrary direction.
Then, for any operator $O$ of finite support, (i.e., $O$ is local,) one has $\langle \psi | O \mathbb{T}^n | \psi \rangle = 0$
whenever $n > 15 w$. The number $15$ is merely a convenient number to make an argument smooth.
Important is that there is some finite $n=n(w)$ such that the transition amplitude becomes
\emph{exactly} zero. See Ref.~\onlinecite{Haah2011Local} for proofs.

The cubic code model is \emph{topologically ordered}~\cite{Wen1991SpinLiquid}
in the sense that the ground state subspace is degenerate
and no local operator is capable of distinguishing any two ground states;~\cite{Haah2011Local}
if $O$ is an arbitrary local operator and $\ket{\psi_1}$ and $\ket{\psi_2}$
are two arbitrary ground states, then one has
\begin{align}
 \langle \psi_1 | O | \psi_1 \rangle = c(O) \langle \psi_1 | \psi_2 \rangle \label{eq:tqo}
\end{align}
for some number $c(O)$ that only depends on the operator $O$ but not on the states $|\psi_{1,2}\rangle$.
In addition, the model satisfies
the so-called ``local topological order'' condition,~\cite{MichalakisZwolak2013Stability}
which implies that the degeneracy is exact up to an error that is exponentially small
in the system size.~\cite{BravyiHastings2011short}
In other words, all ground states have exactly the same local reduced density matrices,
and this property does not require a fine-tuning.
For an application of the model in robust quantum memory, see Ref.~\onlinecite{BravyiHaah2011Memory}.

The actual degeneracy and questions on non-local operators that distinguish
different ground states are fairly technical.
One can show~\cite{Haah2012PauliModule}
that the degeneracy of the cubic code model
on a $L \times L \times L$ lattice with periodic boundary conditions
is equal to $2^k$ where
\begin{align}
\frac{k+2}{4} 
&= \mathrm{deg}_x~ \gcd 
      \begin{bmatrix} 
      1+(1+x)^L,\\
      1+(1+\omega x)^L,  \\
      1+(1+\omega^2 x)^L
      \end{bmatrix}_{\mathbb{F}_4}  \label{eq:degeneracy} \\
&= \begin{cases}
    1 & \text{if $L = 2^p + 1~ (p\ge 1)$},\\
    L & \text{if $L = 2^p ~( p \ge 1 )$}
   \end{cases} \label{eq:degeneracy2}
\end{align}
That is, one computes three polynomials 
over the field of four elements $\mathbb{F}_4 = \{0,1,\omega, \omega^2 \}$
and takes the greatest common divisor polynomial and reads off the degree in $x$.
The proof of this formula contained in Ref.~\onlinecite{Haah2012PauliModule}
is based on an algebraic representation of the Hamiltonian Eq.~\eqref{eq:H-model-A},
which will be reviewed in Sec.~\ref{sec:method} below.

The cubic code Hamiltonian Eq.~\eqref{eq:H-model-A}
belongs to a class of so-called stabilizer (code) Hamiltonians,
as it is defined as a sum of commuting Pauli operators.
The Kitaev toric code model~\cite{Kitaev2003Fault-tolerant}
and the Wen plaquette model~\cite{Wen2003Plaquette}
are well-known examples of stabilizer Hamiltonians.
The ground states in these models have a nice geometric interpretation in terms of
string-nets,~\cite{LevinWen2005String-net}
whereas, unfortunately, there is no known geometric interpretation for the ground state of
the cubic code model, other than the trivial expression Eq.~\eqref{eq:gs}.

\section{Entanglement renormalization and bifurcation}
\label{sec:erg}

It will be useful to recall the notion of \emph{finite depth quantum circuit}.
A depth-1 quantum circuit is a product of local unitary operators of disjoint support.
We do not restrict the number of the unitary operators participating in the product,
but each unitary operator must be local,
that is, its support can be covered by some ball of fixed radius.
This radius is referred to as the range of the circuit.
A finite depth quantum circuit is a finite product of depth-1 quantum circuits.
The number of layers must be independent of system size.
The finite depth quantum circuit is a discrete version 
of the unitary evolution $e^{-it\mathcal H}$ 
by a sum $\mathcal H$ of local Hermitian operators for $t=O(1)$.

The \emph{entanglement renormalization group} transformation
is a procedure where one disentangles some of degrees of freedom
by local unitary transformations,
and compares the transformed state to the original state.
The purpose is to understand ``long range'' entanglement.
Given a many-qubit quantum state $\ket \psi$
and a finite depth quantum circuit $U$ such that 
$U \ket \psi = \ket \phi \otimes \ket{\uparrow} \otimes \cdots \otimes \ket \uparrow$,
we discard the qubits in the trivial state $\ket \uparrow$ from $U \ket \psi$.
Then we proceed with $\ket \phi$ in the next stage of entanglement RG transformations.

The entanglement RG analysis can be done in the Heisenberg picture
when we are interested in a state that is a common eigenstate of a set of operators.
Suppose $\ket \psi$ is defined by equations
\begin{align}
 G_i \ket \psi = \ket \psi \quad \text{for any } i \label{eq:stabilizer}
\end{align}
where $i$ is some index. Then,
the transformed state $U \ket \psi$ is described by equations
\[
(U G_i U^\dagger) U \ket \psi = U \ket \psi .
\]
If $U G_i U^\dagger$ happens to be an operator, say $\sigma^z$ on a single qubit,
then that qubit must be in the state $\ket \uparrow$, disentangled from the others.
This is the criterion by which we identify disentangled qubits
in the calculation below.
In addition, we can use this information to restrict other $G_j$ in the next stage of
entanglement renormalization.

The ground state subspace of our model Eq.~\eqref{eq:H-model-A}
is described by the stabilizer equation~\eqref{eq:stabilizer}
where the stabilizers $G_i$ are just $G^x_i$ and $G^z_i$.
Here, observe that the stabilizers $G_i$ in Eq.~\eqref{eq:stabilizer}
are invertible operators; $G_i$'s form an abelian group $\mathcal G = \langle G_i \rangle$,
called the stabilizer group.
Then, the disentangling criterion is that
for some element $G$ of the stablizer group $\mathcal G$,
$U G U^\dagger $ acts on a single qubit,
where $G$ can be a product of several $G_i$'s.

In fact, only the group $\mathcal G$ is important.
Consider two gapped Hamitonians
\[
 H = -J \sum_i G_i, \quad \quad H' = -J \sum_j G'_j
\]
where the terms $G_i$ and $G'_j$ generate
the same multiplicative group 
$\mathcal G = \langle G_i \rangle = \langle G'_i \rangle$.
The ground-state subspace of the two gapped Hamiltonians are identical
and they represent the same quantum phase of matter,
in which case we will write 
\begin{equation}
H \cong H'.
\label{eq:equivalence}
\end{equation}
One can say that $H'$ is another parent gapped Hamiltonian
of the ground-state subspace of $H$.

Since the ground state is degenerate,
the stabilizer equation~\eqref{eq:stabilizer} 
does not pick out a particular state.
Nevertheless, the disentanglement criterion in the Heisenberg picture
determines a qubit in the trivial state unambiguously for any ground state.
Thus, even after discarding disentangled qubits,
the transformed Hamiltonian $U H U^\dagger$ has a ground-state subspace
that is isomorphic to that of $H$.
Our entanglement RG transformation preserves the ground-state subspace.

We can now state our main result.
Let $H_A(a)$ be the cubic code Hamiltonian defined in Eq.~\eqref{eq:H-model-A}.
Here, the lattice spacing constant $a$ is specified for notational clarity.
We find a finite depth quantum circuit $U$ such that
\begin{align}
U H_A(a) U^\dagger \cong H_A(2a) + H_B(2a) \label{eq:A-to-AB}
\end{align}
where no qubit is involved in both $H_A(2a)$ and $H_B(2a)$.
In Eq.~\eqref{eq:A-to-AB}, 
we have suppressed disentangled qubits; single $\sigma^z$ terms are dropped.
The new model $H_B$ is defined on a simple cubic lattice with four qubits per site,
with the Hamiltonian
\begin{align}
H_B = - J \sum_{i \in \Lambda} \left( S^{x,1}_i + S^{x,2}_i + S^{z,1}_i + S^{z,2}_i \right)
\label{eq:H-model-B}
\end{align} 
where
\begin{align*}
S^{x,1}_i &= \sigma^x_{i+\hat x,1} \sigma^x_{i+\hat z,1} \sigma^x_{i,2} \sigma^x_{i+\hat x,2} 
             \sigma^x_{i+\hat x,3} \sigma^x_{i+\hat y,3} \sigma^x_{i,4} \sigma^x_{i+\hat y,4} \\
S^{x,2}_i &= \sigma^x_{i,1} \sigma^x_{i+\hat x,1} \sigma^x_{i,2} \sigma^x_{i+\hat z,2} 
             \sigma^x_{i,3} \sigma^x_{i+\hat y,3} \sigma^x_{i,4} \sigma^x_{i+\hat x,4} \\
S^{z,1}_i &= \sigma^z_{i,1} \sigma^z_{i-\hat y,1} \sigma^z_{i,2} \sigma^z_{i-\hat x,2} 
             \sigma^z_{i,3} \sigma^z_{i-\hat x,3} \sigma^z_{i,4} \sigma^z_{i-\hat z,4} \\
S^{z,2}_i &= \sigma^z_{i-\hat x,1} \sigma^z_{i-\hat y,1} \sigma^z_{i,2} \sigma^z_{i-\hat y,2} 
             \sigma^z_{i-\hat x,3} \sigma^z_{i-\hat z,3} \sigma^z_{i,4} \sigma^z_{i-\hat x,4}
\end{align*}
are eight-qubit interactions. The interaction terms are visually depicted as
\begin{align*}
\drawgenerator{}{}{XXXI}{}{IIXX}{IXIX}{}{XIII},
\drawgenerator{}{}{XIIX}{}{IIXI}{XXXX}{}{IXII},\\
\drawgenerator{ZIII}{ZZZZ}{}{IIIZ}{}{}{IZZI}{},
\drawgenerator{ZZII}{IZIZ}{}{IIZI}{}{}{ZIZZ}{}.
\end{align*}
We perform a similar entanglement RG transformation for $H_B$,
and find a finite depth quantum circuit $V$ such that
\begin{align}
V H_B(a) V^\dagger \cong H_B(2a) + H_B'(2a) \label{eq:B-to-BB}
\end{align}
with no qubit is involved in both $H_B(2a)$ and $H_B'(2a)$.
$H_B$ and $H_B'$ are the \emph{same} but act on disjoint sets of qubits.
We have dropped single qubits in the disentangled states on the right-hand side of Eq.~\eqref{eq:B-to-BB}.
The proof of these formulas and a compact representation of the models
are given in Sec.~\ref{sec:method}.

The new model $H_B$ is different from the original cubic code model $H_A$.
We will argue in the next section that they represent different quantum phases of matter.
However, they resemble each other in many ways 
because they are related by the finite depth quantum circuit of Eq.~\eqref{eq:A-to-AB}.
Recall that under a finite depth quantum circuit,
any local operator is mapped to a local operator, and the corresponding operator algebras are isomorphic.
In particular, the two models admit pointlike excitations, which are immobile in both cases.
They have degenerate ground states that are locally indistinguishable.

\section{Model A and B are different}
\label{sec:diff-phase}

By the quantum phase of matter, we mean an equivalence class of gapped Hamiltonians
where the equivalence is defined by adiabatic paths in the space of gapped Hamiltonians
and finite depth quantum circuits assisted with some ancillary qubits.~\cite{ChenGuWen2010transformation}
The equivalence may be observed at some different length scale,
so one might have to coarse-grain the lattice in order to see the equivalence.
The nonequivalence, on the other hand, must be proved 
by contrasting some invariants.
We focus on the degeneracy of the ground states for this purpose.

Suppose the two models $A$ and $B$ represent the same quantum phase of matter.
They must have the same ground-state subspace structure,
and in particular the dimension of the ground-state subspace must be the same.
In view of the fluctuating degeneracy as in Eqs.~\eqref{eq:degeneracy},\eqref{eq:degeneracy2},
it means that the degeneracy is given by the same function of the system size under the same boundary conditions.
Let $k_A(L)$ be $\log_2$ of the ground-state subspace dimension of the model $A$, the original cubic code model,
on $L \times L \times L$ periodic lattice,
and let $k_B(L)$ be that of the model $B$, the new model discovered by the entanglement RG transformation.
From Eq.~\eqref{eq:A-to-AB}, we have
\[
k_A(2L) = k_A(L) + k_B(L) .
\]
Eq.~\eqref{eq:degeneracy2} implies that 
\begin{equation}
k_A(2L) = 2k_A(L) +2. \label{eq:kA-doubling}
\end{equation}
Then, it follows that
\begin{equation}
k_B(2L) = 2k_B(L), \label{eq:kB-doubling}
\end{equation}
which can also be shown by Eq.~\eqref{eq:B-to-BB}.
It is clear that the function $L \mapsto k_A(L)$ is \emph{different} from the function $L \mapsto k_B(L)$.
This is the basis of the argument for distinctness of the two phases.

\begin{figure}[tb]
 \includegraphics[width=.45\textwidth]{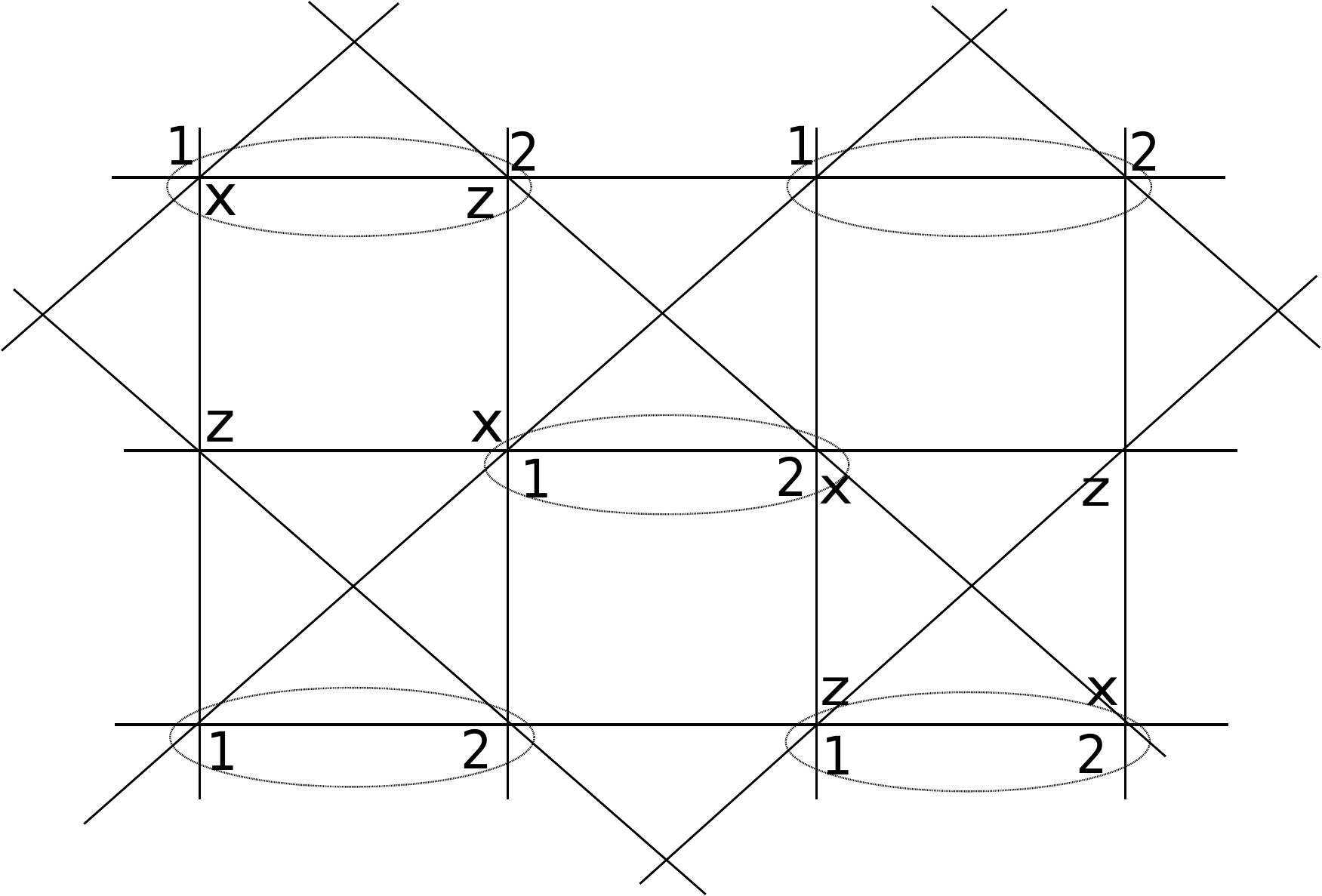}
 \caption{Equivalence between the Wen plaquette model and the toric code model can be observed, only when a unit cell is properly chosen.}
 \label{fig:Wen-to-toric}
\end{figure}

We need to take into account the possibility of the equivalence at different length scales or on distorted lattices.
For example, we know that the Wen plaquette model~\cite{Wen2003Plaquette} exhibits the same phases of matter
as the toric code model.~\cite{Kitaev2003Fault-tolerant}
However, the Wen plaquette model 
\[
 H_\text{Wen} = - \sum_i \sigma^z_{i} \sigma^x_{i + \hat x} \sigma^x_{i+\hat y} \sigma^z_{i+\hat x + \hat y}
\]
has one qubit per lattice site, whereas the toric code model has two.
The degeneracies as functions of system size are different, too.
\[
 k_\text{toric} (L) = 2, \quad k_\text{Wen}(L) = 
 \begin{cases}
  1 & \text{if $L$ is odd,}\\
  2 & \text{if $L$ is even.}
 \end{cases}
\]
To see the equivalence, one has to take a new Bravais lattice for the Wen plaquette model
such that the new unit cell now contains two qubits, and
the unit vectors for the coarser lattice are in the diagonal directions of the original lattice.
The toric code model is recovered once we make local unitary transformations $\sigma^z \leftrightarrow \sigma^x$
on every, say, first qubit in each new unit cell. See Fig.~\ref{fig:Wen-to-toric}.

For the most general choice of new Bravais lattice (smaller translation group) in the cubic lattice,
the new unit translation vectors have integer coordinates such that
the $3 \times 3$ matrix $M$ of the cooridnates in the columns is nonsingular.
The unit vectors define a rhombohedron unit cell.
Conversely, given a $3 \times 3$ nonsingular integer matrix $M$,
one can introduce a new Bravais lattice to the original cubic lattice
by declaring the columns of $M$ to be new unit translation vectors.
Imposing periodic boundary conditions amounts to specifying 
the number of translations in each new direction $\vec L' = (L_x', L_y', L_z')$
before the translations become the identity translation.
Hence, the degeneracy under periodic boundary conditions is a function of $M$ and the lattice dimension vector $\vec L'$;
$k = k(M, \vec L')$.

Suppose now that two models $H_A$ and $H_B$ are equivalent,
and the equivalence is made explicit at coarser lattices $\Lambda'_A$ and $\Lambda'_B$
defined by nonsingular integer matrices $M_A$ and $M_B$, respectively, with respect to the original cubic lattice $\Lambda$.
In particular, we must have
\[
k_A \left( M_A, \vec{L'} \right) = k_B \left( M_B, \vec{L'} \right) 
\]
for any lattice dimension vector $\vec{L'}$.
Consider an even coarser lattice $\Lambda''_A$ defined by a nonsingular integer matrix $N$ with respect to $\Lambda'_A$,
and $\Lambda''_B$ defined by the same $N$ with respect to $\Lambda'_B$.
We must have
\[
 k_A \left( M_A N, \vec{L''} \right) = k_B \left( M_B N, \vec{L''} \right) 
\]
for any lattice dimension vector $\vec{L''}$.
Note that $N$ was arbitrary.

Set the matrix $N$ to be the adjugate matrix of $M_A$ so that $M_A N = \det (M_A) I_{3 \times 3}$.
$N$ is nonsingular and integral.
For $\vec{L''} = (\ell, \ell, \ell)$, we have
\begin{align*}
 k_A \left( \det (M_A) I_{3 \times 3}, (\ell, \ell, \ell) \right) 
 &= k_B \left( M_B N, (\ell, \ell, \ell) \right) \\
 &= k_A \left( \det(M_A) \ell \right),
\end{align*}
where the last function is one that has appeared in Eq.~\eqref{eq:kA-doubling}.
The function $\phi_B: \ell \mapsto k_B \left( M_B N, (\ell, \ell, \ell) \right)$ has a property
that $\phi_B(2\ell) = 2 \phi_B(\ell)$ because of Eq.~\eqref{eq:B-to-BB}, regardless of how $M_B$ or $N$ is chosen.
However, we know from Eq.~\eqref{eq:kA-doubling} that the function $\phi_A : \ell \mapsto k_A \left( \det(M_A) \ell \right)$
has a property that $\phi_A(2\ell) = 2 \phi_A(\ell) +2$.
This is a contradiction, and therefore the model $H_A$ and $H_B$ represents different phases of matter.

\section{A tensor network description: Branching MERA}
\label{sec:bMERA}

The entanglement RG transformation yields
a tensor network description for the state.
If one reverses the transformation
starting from, say, a state on $L^3$ lattice,
one gets a state on $(2L)^3$ lattice.
After many iterations one obtains a state on an
infinite lattice. It will be an exact description
since our finite depth quantum circuits $U, V$ are exact.
In this section we will refer to local degrees of freedom as qudits.

Let us review Multi-scale Entanglement Renormalization Ansatz (MERA) states.~\cite{Vidal2007ER,Vidal2008MERA}
The MERA state is a many-qudit state that is obtained 
by reversing the entanglement RG transformations as follows.
One starts with a qudit system on some lattice.
(Step 1) Apply a finite depth quantum circuit with 
some ancillary qudits in a fixed state $\ket \uparrow$.
Due to the insertion of the ancillary qudits,
the number density of qudits is increased.
In order to retain the number density,
(Step 2) one expands the lattice.
Then, (Step 3) Iterate Step 1 and 2.
In a scale invariant system,
one expects that the quantum circuit in Step 1
is the same for every level of the iterations.
The class of states that can be written as a MERA
is proposed to describe ground states of some critical systems,
and is shown to admit efficient classical algorithms.

Since the ground state of the toric code model for example is an entanglement RG
fixed point, it naturally has a scale-invariant MERA description.
On the other hand, the cubic code model is not a usual fixed point.
At a coarse-grained level,
the ground-state subspace is a tensor product of two independent
ground-state subspaces (Eq.~\eqref{eq:A-to-AB}),
each of which is again a tensor product of two independent ground-state
subspaces (Eqs.~\eqref{eq:A-to-AB},\eqref{eq:B-to-BB}).
Reversing the entanglement RG flow,
we see that the final state is obtained by entangling
two states, each of which is again obtained by
entangling two states, and so on.

The ``branching MERA,'' recently introduced by Evenbly and Vidal,~\cite{EvenblyVidal2012RG}
is a variant of MERA that captures this scenario.
In a branching MERA, the ancillary trivial qudits  in the Step 1 of the usual MERA
are allowed to be in branching MERA states.
The self-referential nature is essential.
The total number of branches would grow exponentially with the coarse-graining level.

The branching structure usually yields very highly entangled states.
For example, in a 1D spin chain, a typical branching MERA state with 
the total number of branches being $b_n = 2^n$ at coarse-graining level $n$,
obeys a ``volume'' law of entanglement entropy.
In general, the entanglement entropy of a ball-like region of linear dimension $L$,
for a branching MERA state in a $D$-dimensional lattice scales like
\begin{equation}
 S \le O(1) \sum^{\log_2 L}_{n=0} b_n \left(\frac{L}{2^n}\right)^{D-1} \label{eq:bMERA-ent-scaling}
\end{equation}
where $b_n$ is the total number of branches at RG level $n$.~\cite{EvenblyVidal2013bMERAentropy}
A proof of the formula is given in Appendix~\ref{sec:entropy-bound}.
In case of our cubic code model where $b_n = 2^n$, it gives an area law.
It is consistent with the fact that it is a stabilizer code Hamiltonian.~\cite{HammaIonicioiuZanardi2005}

It should be noted that the entanglement entropy scaling alone 
does not necessitate the branching structure;
it does not nullify the possibility of a description by the usual unbranched MERA.
Our bound in Eq.~\eqref{eq:bMERA-ent-scaling} merely illustrates that
the branching MERA description of the cubic code model
is consistent in view of the entanglement entropy scaling,
despite the intuition that the branching MERA yields much more entanglement.

Rather, the necessity of the branching structure relies on the ground state degeneracy.
If a usual MERA description were possible, the ground space of the cubic code model on $L^3$ (with $L=2^n$) lattice
would have a one-to-one correspondence with the Hilbert space of $O(1)=O(L^0)$ qubits
in the top level of the MERA, and therefore would be of a constant dimension.
This would contradict Eq.~\eqref{eq:degeneracy2}.

\section{Calculation method}
\label{sec:method}

The finite depth quantum circuits $U$ and $V$
are complicated and not very enlightening.
Explicit circuits and calculation can be found in
a Mathematica script in Supplementary Material.~\cite{SM}
In this section, we explain a machinary to compute $U$ and $V$.
It heavily depends on a special structure of the Hamiltonians $H_A$
and $H_B$.
The content here is essentially presented in Ref.~\onlinecite{Haah2012PauliModule},
so we will be brief.

\subsection{Laurent polynomial matrix description}

The Pauli $2 \times 2$ matrices $\sigma^x, \sigma^y, \sigma^z$ 
have a special property that (i) they square to identity,
(ii) the product of any pair of the matrices 
results in the third up to a phase factor $\pm 1, \pm i$,
and (iii) they anticommute with one another.
In other words, they form an abelian group under multiplication
up to the phase factors. This group, ignoring the phase factors,
is just $\mathbb{Z}_2 \times \mathbb{Z}_2$.
A conventional correspondence is given by
\begin{align}
 (\sigma^x)^n (\sigma^z)^m 
 & \in  \langle \sigma^x, \sigma^y, \sigma^z \rangle / \{ \pm 1, \pm i\} \nonumber \\
 \Updownarrow& &\\
  (n,m) &\in \mathbb{Z}_2 \times \mathbb{Z}_2 \nonumber
\end{align}
The correspondence easily generalizes to Pauli operators
(tensor products of Pauli matrices).
An $n$-qubit Pauli operator corresponds to a bit $\{0,1\}$ string of length $2n$:
The first half of the bit string expresses $\sigma^x$,
while the second half expresses $\sigma^z$.

If a qubit system admits translations, e.g.~one-dimensional spin chain,
the corresponding bit string can be written in a compact way:
Write the bits in the coefficients of the translation group elements
in a formal linear combination. For example,
\begin{align}
\cdots\otimes \sigma^x \otimes \sigma^z \otimes I \otimes \sigma^y \otimes \cdots \nonumber \\
\Leftrightarrow \begin{pmatrix}
\cdots & 1 & 0 & 0 & 1 & \cdots \\
\cdots & 0 & 1 & 0 & 1 & \cdots
\end{pmatrix} \label{eq:poly-mapping}\\
\Leftrightarrow \begin{pmatrix}
\cdots + 1 t^{-1} + 0 t^0 + 0 t^1 + 1 t^2 + \cdots \\
\cdots + 0 t^{-1} + 1 t^0 + 0 t^1 + 1 t^2 + \cdots
\end{pmatrix} \nonumber \\
= \begin{pmatrix}
\cdots + t^{-1} + t^2 + \cdots \\
\cdots + 1 + t^2 + \cdots
\end{pmatrix} \nonumber
\end{align}
where $t$ denotes the translation by one unit length to the right.
This is merely a change of notation.
It yields a particularly simple expression
for translation-invariant Hamiltonians whose terms are Pauli operators,
because one only has to keep a few polynomials that express different types of local terms.
Local terms are expressed not by an infinite Laurent series, 
but by a finite linear combination of the translation group elements.
Summarizing, we have introduced a notation for Hamiltonians of Pauli operators
using the translation group algebra with coefficients in $\mathbb{Z}_2$.

The cubic code model $H_A$ in Eq.~\eqref{eq:H-model-A} can now be written as
\begin{align}
G^x = \begin{pmatrix}
1+x+y+z \\
1+xy+yz+zx \\
0\\
0
\end{pmatrix},~~
G^z = \begin{pmatrix}
0\\
0\\
1+\bar x \bar y + \bar y \bar z + \bar z \bar x \\
1+ \bar x + \bar y + \bar z
\end{pmatrix}.
\end{align}
where $x,y,z$ are translations along 
$+\hat x,+\hat y,+\hat z$-direction, respectively,
and $\bar x = x^{-1}$, etc.
Since the unit cell of the cubic code model contains
two qubits, we need $2 \times 2 = 4$ rows in the matrix.
The first row expresses $\sigma^x$ in the first qubit
at each site, the second row $\sigma^x$ the second qubit,
the third row $\sigma^z$ in the first qubit,
and the fourth row $\sigma^z$ in the second qubit.
It is the most convenient to write two matrices in a single matrix
where each type of term is written in each column.
\begin{align}
\sigma = \begin{pmatrix}
1+x+y+z  & 0 \\
1+xy+yz+zx & 0\\
0& 1+\bar x \bar y + \bar y \bar z + \bar z \bar x\\
0&1+ \bar x + \bar y + \bar z
\end{pmatrix} \label{eq:matrix-sigma-cubic}
\end{align}
We refer to this matrix $\sigma$ as a \emph{generating matrix}
of $H_A$.

\subsection{Applying periodic local unitary operators}

A subclass of finite depth quantum circuits is effectively
implemented using this Laurent polynomial description.
It consists of unitaries that respect the translation symmetry
and map Pauli operators to Pauli operators.
More specifically, they are compositions of so-called CNOT,
Hadamard, and Phase gates.
For example, Hadamard gate 
\[
U_\mathrm{Hadamard} = \frac{1}{\sqrt{2}}\begin{pmatrix} 1 & 1 \\ 1 & -1 \end{pmatrix}
\begin{matrix} \ket \uparrow \\ \ket \downarrow \end{matrix}
\]
swaps $\sigma^x$ and $\sigma^z$:
\[
U_H \sigma^x U_H^\dagger = \sigma^z, ~~ U_H \sigma^z U_H^\dagger = \sigma^x
\]
If the Hadamard is applied for every qubit on the lattice,
then the upper half and the lower half of the Laurent polynomial matrix
will be interchanged.
Similarly, one can work out the action of the CNOT gate
\[
U_\mathrm{CNOT} =
\begin{pmatrix}
  1 & 0 & 0 & 0 \\
  0 & 1 & 0 & 0 \\
  0 & 0 & 0 & 1 \\
  0 & 0 & 1 & 0
\end{pmatrix}
\begin{matrix}
 \ket{ \uparrow \uparrow }\\
 \ket{ \uparrow \downarrow }\\
 \ket{ \downarrow \uparrow }\\
 \ket{ \downarrow \downarrow }
\end{matrix}
\]
and Phase gate
\[
U_\mathrm{Phase} =
\begin{pmatrix}
 1 & 0 \\
 0 & i 
\end{pmatrix}
\begin{matrix} \ket \uparrow \\ \ket \downarrow \end{matrix}
\]
on the Laurent polynomial matrix.
The result is that they correspond to \emph{row operations}
on the Laurent polynomial matrix.
That is, any elementary row operation $E$, viewed as a left matrix multiplication
$\sigma \mapsto E \sigma$,
is admissible as long as $E$ satisfies the symplectic condition
\begin{equation}
\bar E^T \begin{pmatrix} 0 & I_q \\ I_q & 0 \end{pmatrix} E = 
\begin{pmatrix} 0 & I_q \\ I_q & 0 \end{pmatrix} \mod 2.
\label{eq:symplectic}
\end{equation}
where the bar means the antipode map under which $x \mapsto x^{-1}$,
$y \mapsto y^{-1}$, and $z \mapsto z^{-1}$.
Here, $q$ is the number of qubits per unit cell.
$I_q$ is the $q \times q$ identity matrix.
For a proof, see Ref.~\onlinecite{Haah2012PauliModule}.

Note that when the two-qubit unitary operator CNOT above
acts within a unit cell, the antipode map is trivial 
since $E$ in Eq.~\eqref{eq:symplectic} will not involve any variable $x,y,z$, etc;
the antipode map does not do anything to coefficients.
When the CNOT acts on a pair of qubits across the unit cells,
which is allowed only if the unit cell contains two or more qubits,
the antipode map is nontrivial.
Of course, in any case, the overall unitary must have the same periodicity with the lattice.

Using the above row operations, one can only generate
a finite depth quantum circuit whose periodicity is $1$.
If one wishes to apply, say, Hadamard gates on every other qubits (periodicity 2),
one has to choose a subgroup $\mathcal T'$ of 
the original translation group $\mathcal T$,
so that one unit of translation under $\mathcal T'$ is the translation
by two units under $\mathcal T$.
Then, one can implement the periodicity $2$ quantum circuit,
using the prescription above.
Under such a coarse translation group, our matrix representation
of the Hamiltonian must be different.
Computing a new representation is easy,
and a prescription is as follows.
If one wishes to take the coarse translation group to be 
\[
\mathcal T' = \langle x', y, z \rangle \le \langle x,y,z \rangle = \mathcal T
\]
where $x' = x^2$,
one simply replaces each Laurent polynomial $f(x,y,z)$ of $\sigma$ 
with the matrix
\begin{align}
f \left( \begin{pmatrix} 0 & x'\\ 1 & 0 \end{pmatrix}, \begin{pmatrix}y & 0 \\ 0 & y \end{pmatrix}, \begin{pmatrix}z & 0 \\ 0 & z \end{pmatrix} \right)
\label{eq:coarse-grain-prescription}
\end{align}
If the old generating matrix $\sigma$ was $2q \times m$,
then the new generating matrix is $4q \times 2m$.
Again, a proof of this claim can be found in Ref.~\onlinecite{Haah2012PauliModule}.

\subsection{Example: Toric code model}
\label{sec:egRG-toric-code}

Let us perform an entanglement RG for the toric code model
(Ising gauge theory).~\cite{Kitaev2003Fault-tolerant}
As we call for strict translation-invariance,
we take the square lattice with the unit cell at a vertex consisting of
one horizontal edge on the east (1) and one vertical edge on the north (2).
The Hamiltonian is
\begin{align*}
H_\mathrm{toric} = &- \sum_i \sigma^x_{i,1}
     \sigma^x_{i-\hat x,1} \sigma^x_{i,2} \sigma^x_{i-\hat y,2} \\
 &- \sum_i \sigma^z_{i,1} \sigma^z_{i+\hat y,1}
    \sigma^z_{i,2} \sigma^z_{i+\hat x,2}
\end{align*}
Following the correspondence Eq.~\eqref{eq:poly-mapping},
the generating matrix is
\begin{align}
\sigma_\mathrm{toric} = \begin{pmatrix}
1+\bar x & 0 \\
1+\bar y & 0 \\
0 & 1 + y \\
0 & 1 + x
\end{pmatrix}. \label{eq:matrix-sigma-toric}
\end{align}
Let us take a smaller translation group $\mathcal T' = \langle x' ,y \rangle \le \langle x,y\rangle$ where $x' = x^2$.
Accordingto the prescription Eq.~\eqref{eq:coarse-grain-prescription},
the new generating matrix with respect to $\mathcal T'$ becomes
\begin{align}
\sigma'_\mathrm{toric} =
\begin{pmatrix}
1       & 1 &  &  \\
\bar x' & 1 &  &  \\
1+\bar y& 0 &  &  \\
0       & 1+\bar y &  & \\
 &  & 1+y & 0 \\
 &  & 0   & 1+y \\
 &  & 1 & x' \\
 &  & 1 & 1  
\end{pmatrix} \label{eq:coarse-grained-toric-code}
\end{align}
Some zeros are not shown.
Now we apply row operations that satisfy Eq.~\eqref{eq:symplectic}.
\begin{align}
\left(
\begin{array}{cccccccc}
 1 & 0 & 0 & 0 &  &  &  &  \\
 \bar x' & 1 & 0 & 0 &  &  &  &  \\
 \bar y+1 & 0 & 1 & 0 &  &  &  &  \\
 \bar y+1 & 0 & 1 & 1 &  &  &  &  \\
  &  &  &  & 1 & x' & 1+y & 0 \\
  &  &  &  & 0 & 1 & 0 & 0 \\
  &  &  &  & 0 & 0 & 1 & 1 \\
  &  &  &  & 0 & 0 & 0 & 1 \\
\end{array}
\right)\sigma'_\mathrm{toric}\nonumber \\
= 
\left(
\begin{array}{cccc}
 1 & 1 &  &  \\
 0 & \bar x'+1 &  &  \\
 0 & \bar y+1 &  &  \\
 0 & 0 &  &  \\
  &  & 0 & 0 \\
  &  & 0 & 1+y \\
  &  & 0 & 1+x' \\
  &  & 1 & 1 \\
\end{array}
\right) \label{eq:toric-rg}
\end{align}
Let us recover the Hamiltonian.
We have found a finite depth quantum circuit $U$ from Eq.~\eqref{eq:toric-rg}
such that
\begin{align*}
&U H_\mathrm{toric} U^\dagger\\
&= - \sum_{i'} \sigma^x_{i',1} - \sum_{i'} \sigma^x_{i',1} \sigma^x_{i',2} \sigma^x_{i'-\hat x',2} \sigma^x_{i',3} \sigma^x_{i'-\hat y,3} \\
& - \sum_{i'} \sigma^z_{i',4} - \sum_{i'} \sigma^z_{i',2} \sigma^z_{i'+\hat y,2} \sigma^z_{i',3} \sigma^z_{i'+\hat x',3} \sigma^z_{i',4} .
\end{align*}
Since the Hamiltonian is frustration-free,
it is clear that the first and fourth qubits in each unit cell
are in a trivial state and are disentangled from the rest.
As noted above in Sec.~\ref{sec:erg},
only the multiplicative group generated by the terms
in the Hamiltonian is important,
and we recover $H_\mathrm{toric}$ we started with
at a coarse-grained lattice $\mathcal T'$.
The example demonstrates that \emph{any column operation on the generating matrix $\sigma$ is allowed}
in view of equivalence Eq.~\eqref{eq:equivalence}.
This shows that the ground state of the toric code model
is a fixed point in an entanglement RG flow.~\cite{AguadoVidal2007Entanglement}

In Supplementary Material,
we perform similar calculations for 3D and 4D toric code models.
(3D toric code model is also known as 3D Ising gauge theory.~\cite{Wegner1971IsingGauge}
4D toric code is similar; qubits live on plaquettes,
and the gauge transformation flips qubits around an edge.~\cite{DennisKitaevLandahlEtAl2002Topological})
We verify that they are all entanglement RG fixed points.

\section{An algebro-geometric test on entanglement RG}
\label{sec:alg-check}

Our example of the bifurcation is very specific to the cubic code model,
and general criteria for the bifurcation to happen are not well understood.
However, we can rule out certain possibilities as follows.
We have found an equivalence by a finite depth quantum circuit 
between the ground space of $H_A(a)$, 
where $a$ in the parentheses is the lattice spacing,
and that of $H_A(2a) \oplus H_B(2a)$.
Can we find a similar relation between
the ground space of $H_A(a)$ and that of, say, $H_A(3a) \oplus H'$ for some Hamiltonian $H'$?
Put differently, how coarse should a new Bravais lattice be,
if one wishes to find a copy of $H_A$ on the new Bravais lattice
by a finite depth quantum circuit?

In this section, we give a \emph{necessary} condition for this question to be answered positively
by exploiting our Laurent polynomial matrix descriptions.
The condition will detect cases 
when one will not find a copy of the original model one started with on a coarser lattice.
Our choice of new Bravais lattice of lattice spacing $2a$ 
for the cubic code model and the toric code model
satisfies the condition, as it must do.

Let us restrict ourselves to the simplest situation
where the generating matrix $\sigma$ is $2q \times q$, where $q$ is even, and block-diagonal,
as in Eq.~\eqref{eq:matrix-sigma-cubic} and Eq.~\eqref{eq:matrix-sigma-toric}.
This is the case when the number of qubits in the unit cell is the same as
the number of interaction types in the Hamiltonian.
Note that in either Eq.~\eqref{eq:matrix-sigma-cubic} or Eq.~\eqref{eq:matrix-sigma-toric},
the upper-left block is described by two polynomials $f,g$:
For the cubic code model, they are $1+x+y+z$ and $1+xy + yz+ zx$.
For the toric code model, they are $1+x^{-1}$ and $1+y^{-1}$.
The lower-right blocks in both cases are related to the upper-left blocks by the antipode map,
so we can focus only on the upper-left blocks.

Consider all $q/2 \times q/2$ submatrices of the upper-left block of the generating matrix $\sigma$,
and take the determinants of them.
Let $I(\sigma) = \{ f_i \}$ be the set of all such determinants.
For example, $I(\sigma_\text{toric}) = \{ 1+x^{-1}, 1+y^{-1} \}$, and $I(\sigma_\text{cubic}) = \{ 1+x+y+z, 1+xy+yz+zx\}$.
Let $V(\sigma)$ be the set of solutions of the polynomial equations $f_i = 0$.
For example, $V(\sigma_\text{toric}) = \{ (x,y) | 1+ x^{-1} = 0,~ 1+ y^{-1} = 0 \} = \{(1,1)\}$.
It is shown in Ref.~\onlinecite{Haah2012PauliModule} that 
$V(\sigma)$ is invariant under a class of local unitary transformations
such that the transformed Hamiltonian still admits a description by a Laurent polynomial matrix.
$V(\sigma)$ is the object for our algebro-geometric test.

$V(\sigma)$ is a variety, a rather abstract geometric set.
In our Laurent polynomial matrix description,
the variables $x,y$, etc. were directly related to translations.
But, now we are treating them as unknown variables
and furthermore equating the polynomials in those variables with zero!
Indeed, it requires good deal of preparation before defining the variety properly,
which is out of the scope of the present paper.
We will state facts that are useful for our purpose.
Interested readers are referred to Ref.~\onlinecite{Haah2012PauliModule}.

We have seen in Sec.~\ref{sec:egRG-toric-code} that the generating matrix $\sigma$
takes a different form $\sigma \to \sigma'$ depending on our choice of translation group.
Upon taking a coarse translation group, the variety is changed to $V(\sigma) \to V(\sigma')$.
Interestingly, one can show that the change is again given by a nice algebraic map.
For example, if we take 
\[
 \mathcal T' = \langle x', y', z' \rangle \le \langle x,y,z \rangle = \mathcal T
\]
where $x' = x^n$, $y' = y^n$, and $z' = z^n$ in three-dimensional lattice,
which means $n^3$ sites are blocked to form a single new site,
then the change is given by an almost surjective map%
\footnote{Rigorously speaking, the image of the map is dense in the target variety under Zariski topology.
See e.g. Hartshorne, {\it Algebraic Geometry}, Springer}
\begin{equation}
 V(\sigma) \ni (a,b,c) \mapsto (a^n, b^n, c^n) \in V(\sigma'). \label{eq:powering-map}
\end{equation}

The variety $V(\sigma_1 \oplus \sigma_2)$ for the juxtaposition 
of two independent systems $\sigma_1$ and $\sigma_2$ as in Eq.~\eqref{eq:A-to-AB},
is given by the union $V(\sigma_1) \cup V(\sigma_2)$ of respective varieties.

We have noted that $V(\sigma)$ is invariant under local unitary transformations.
The entanglement RG is a combination of local unitary transformations after a choice of a smaller translation group.
Hence, if a copy of the original model is to be found in the coarse lattice,
\emph{the new variety $V(\sigma')$ must contain the original $V(\sigma)$.}
This is a criterion by which the bifurcation, 
or an occurrence of the original model at a coarse lattice \emph{may} happen.
It is unknown if the criterion is a sufficient condition.

\subsection{Examples}

Let us apply the criterion to the toric code model and the cubic code model.
As we have seen above, $V(\sigma_\text{toric}) = \{ (1,1) \}$.
Upon a choice of a coarser lattice, blocking $2 \times 2$ sites as a new one site,
the variety is transformed by the map $ x \mapsto x^2$ and $y \mapsto y^2$.
Obviously, the point $(1,1)$ is invariant under this map, which is consistent with the fact that the toric code is a RG fixed point.~\cite{AguadoVidal2007Entanglement}
(See Sec.~\ref{sec:egRG-toric-code}.)
The readers are encouraged to compute $V(\sigma'_\text{toric})$ from Eq.~\eqref{eq:coarse-grained-toric-code}
and Eq.~\eqref{eq:toric-rg}:
Compute the determinants of all possible $2 \times 2$ submatrices of the upper-left block of $\sigma'_\text{toric}$,
equate them with zero, and decide the set of solutions.

For the cubic code, the variety is also simple. 
It consists of two lines each of which is parametrized by an auxiliary variable $s$:
\begin{align*}
\begin{cases}
 x &= 1+ s \\
 y &= 1+ \omega s \\
 z &= 1+ \omega^2 s
\end{cases},
\quad
\begin{cases}
 x &= 1+ s \\
 y &= 1+ \omega^2 s \\
 z &= 1+ \omega s
 \end{cases}.
\end{align*}
where $\omega$ is a third root of unity satisfying $\omega^2 + \omega + 1 = 0$.
(It should be noted that the numbers are not complex numbers;
they belong to extension fields of the binary field $\mathbb{F}_2$.)
On a coarser lattice blocking $2^3$ sites together,
the variety is transformed by the squaring map. See Eq.~\eqref{eq:powering-map}.
Over the binary field, $(a+b)^2 = a^2 + 2ab + b^2 = a^2 + b^2$ for any $a,b$.
Hence, the image of the squaring map is the union of two lines
\begin{align*}
\begin{cases}
 x &= 1+ s^2 \\
 y &= 1+ \omega^2 s^2 \\
 z &= 1+ \omega s^2
\end{cases} \quad
\begin{cases}
 x &= 1+ s^2 \\
 y &= 1+ \omega s^2 \\
 z &= 1+ \omega^2 s^2
\end{cases}
\end{align*}
This is indeed the original variety, although the two lines are interchanged by the squaring map.
This is consistent with the fact that we have found the original copy $H_A$ in the coarse lattice.

Note that the varieties for $H_A$ and $H_B$ are the same.
They do not distinguish two different phases of matter;
the variety is a crude algebro-geometric object associated to the Hamiltonian.

\begin{figure}[tb]
 \includegraphics[width=.35\textwidth]{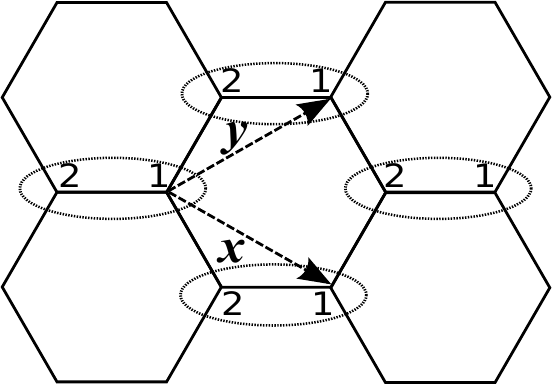}
 \caption{Honeycomb lattice with qubits numbered within a unit cell.}
 \label{fig:honeycomb-color-code}
\end{figure}

Before concluding the section,
we illustrate an example where the test helps to choose a correct new unit cell.
The color code model,~\cite{BombinMartinDelgado2006ColorCode}
which is known to be equivalent to two copies of the toric code model,~\cite{BombinDuclosCianciPoulin2012}
lives on a honeycomb lattice with one qubit at each vertex.
Being a hexagon, any plaquette $p$ has six vertices $v$.
The color code model is defined by the Hamiltonian
\[
 H = - J \sum_{p} \left( \prod_{v \in p} \sigma^z_v + \prod_{v \in p} \sigma^x_v \right) ,
\]
where the sum is over all hexagons.
This is expressed with Pauli matrices and each term commutes with any other,
and thus our Laurent polynomial matrix description is applicable.
Since the honeycomb lattice has two vertices in the conventional unit cell (Fig.~\ref{fig:honeycomb-color-code}),
our generating matrix $\sigma_\text{color}$ is $4 \times 2$, as in the toric code model.
Explicitly,
\[
 \sigma_\text{color} = 
 \begin{pmatrix}
  1+x+y & 0 \\
  x+y+xy & 0 \\
   0         & 1+x+y \\
   0         & x+y+xy
 \end{pmatrix}.
\]
The associated variety is
\begin{align*}
 V(\sigma_\text{color}) &= \{ (x,y) ~|~1+x+y = 0,~ x+y+xy=0 \} \\
 &= \{ (\omega, \omega^2), ~(\omega^2, \omega) \},
\end{align*}
where $\omega$ is a third root of unity over the binary field.

Suppose one tries to find a copy of itself at a coarser lattice,
to see if the model is an entanglement RG fixed point.
One could choose a new Bravais lattice $\Lambda'$ by saying that $x' = x^3$ and $y' = y^3$
are new unit translations. According to Eq.~\eqref{eq:powering-map},
the new variety $V(\sigma'_\text{color})$
would be a single point $(1,1)$ since $\omega^3 = (\omega^2)^3= 1$.
The original variety is not contained in the new variety,
and therefore one will not find a copy of the original model on the coarse Bravais lattice $\Lambda'$.

On the other hand, if one tried to show the equivalence of the color code model
and the toric code model, then one should take the mentioned Bravais lattice $\Lambda'$;
otherwise, the variety of the transformed color code model
would not match that of the toric code model,
and the equivalence would never be explicit.

\section{Discussion}
\label{sec:discussion}

We have shown that under the entanglement renormalization group flow
the cubic code model bifurcates.
The cubic code model $A$ does not simply produce exactly the same two copies of itself,
but yields a different model $B$.
In order to complete the entanglement RG, 
we have further shown that the model $B$ bifurcates into two copies of itself.

The bifurcation alone, as seen in phase B, can be observed in a trivial and rather ad hoc example:
An infinite stack of toric codes. We need to be a little formal because the example is too trivial.
Let $H_\text{toric}(a)$ be the Hamiltonian of the toric code model on a 2D square lattice with qubits on edges,
where lattice spacing is $a$.
The entanglement RG transformation reveals that there is a finite depth quantum circuit $U$ such that
\[
 U H_\text{toric}(a) U^\dagger \cong H_\text{toric}(2a)
\]
Consider an infinite stack of 2D square lattices with qubits on the edges.
Suppose each layer is parallel to $xy$-plane, and the total system is stacked in $z$-direction.
Our ad hoc Hamiltonian is
\[
 H_\text{stack}(a) = \sum_{z = -\infty}^\infty H_\text{toric}(a)_z ,
\]
where the subscript $z$ designates the layer that $H_\text{toric}(a)$ lives on.
Choosing a new Bravais lattice such that $(0,0,2)$ is a new unit translation vector,
we have
\[
 H_\text{stack}(a) = \sum_{z'=-\infty}^\infty H_\text{toric}(a)_{2z'} + H_\text{toric}(a)_{2z'+1} .
\]
Let $V = \bigotimes_{z = -\infty}^\infty U_z$ be a finite depth quantum circuit where $U_z$ is just $U$ acting on the layer $z$.
Then,
\begin{align*}
 V H_\text{stack}(a) V^\dagger &= \sum_{z'=-\infty}^\infty U_{2z'} H_\text{toric}(a)_{2z'} U_{2z'}^\dagger \\
 &~~ +  \sum_{z'=-\infty}^\infty U_{2z'+1} H_\text{toric}(a)_{2z'+1} U_{2z'+1}^\dagger \\
 &\cong \sum_{z'=-\infty}^\infty H_\text{toric}(2a)_{2z'} \\
 &~~+ \sum_{z'=-\infty}^\infty H_\text{toric}(2a)_{2z'+1}\\
 &= H_\text{stack}(2a)_\text{even} + H_\text{stack}(2a)_\text{odd}.
\end{align*}

In contrast, our model cannot be written as a stack of lower dimensional systems.
If it were possible, the ground state degeneracy could not have such complicated dependence on the system size;
at least one parameter, say $L_z$ must be factored out from Eq.~\eqref{eq:degeneracy}.
The fact that the model A and the model B are different 
gives a more direct proof that the model $A$ cannot be described in terms of 2D systems.
If the model $A$ was a stack of lower dimensional ones, the entanglement RG would have
yielded the same two copies of itself.

In our tensor network description, the branching MERA,
one parametrizes states by a network of tensors.
The topology of the network is fixed and the entanglement RG changes
the values of components of the tensors ---
It is the space of tensors where the entanglement RG flows.
It should be pointed out, however, that in our calculation of entanglement RG
the disentangling transformations are obtained accidentally.
The calculation was not guided by any equation,
but we just tried to disentangle as many qubits as possible
and discovered that the state belongs to the ground space 
of two independent systems.
(In fact, the only guide was the consistent behavior of the algebraic variety
under a choice of a new Bravais lattice.)
This motivates us to establish RG equations
that incorporates the branching structure.
In previous studies in this direction,~\cite{VerstraeteCiracLatorreEtAl2005Renormalization,GuLevinWen2008Tensor-entanglement}
it was implicitly assumed that there is no branching at the coarse-grained level.

Recently, Swingle~\cite{Swingle2013} has shown several examples where entanglement entropy does not decrease
under renormalization group transformations, and argued that the so-called $c$-theorem~\cite{Zamolodchikov1986cthm}
and its higher dimensional analogs~\cite{JafferisKlebanovPufuEtAl2011,KomargodskiSchwimmer2011}
can be violated if Lorentz symmetry is broken.
In other words, he argues that the entanglement entropy
is not a RG-monotone in non-Lorentz-invariant theories.
Our example is a yet different (counter)example to those RG-monotone theorems.
The picture that the number density of effective degrees of freedom
should decrease under RG, is manifestly broken.
Although it is not straightforward to directly relate our entanglement RG and the field-theoretic RG,
it will not be the case that in any renormalizable field theory
the number of distinct fields increases as the probing energy scale decreases.
This suggests that the model admits no conventional field theory description
that gives the correct ground space.

\begin{acknowledgments}
The author would like to thank Guifre Vidal for raising a question that has resulted in this work.
The author also thanks John Preskill and Glen Evenbly for numerous helpful discussions.
A part of this work was done at IBM Watson Research Center, Yorktown Heights, New York,
where the author was a summer research intern.
The author is supported in part by Caltech Institute for Quantum Information and Matter,
an NSF Physics Frontier Center with support from Gordon and Betty Moore Foundation,
and by MIT Pappalardo Fellowship in Physics.
\end{acknowledgments}

\appendix

\section{Entanglement entropy of branching MERA states}
\label{sec:entropy-bound}

In this section, we bound the entanglement entropy of a branching MERA~\cite{EvenblyVidal2012RG} state
between some ball-like region and its complement by a function of the region's size.
The proof here will be a simplified version of Ref.~\onlinecite{EvenblyVidal2013bMERAentropy}.
We will relate the entropy scaling with spatial dimension and the number of branches.
A simple lemma will be useful.
Each qudit has Hilbert space dimension $\chi$.

{\bf Lemma.} Let $A,B,C,D$ be disjoint sets of qudits of dimension $\chi$,
and $U$ be a unitary operator acting on $B$ and $C$.
Let $S_{AB}(\rho) = S( \tr_{(AB)^c} \rho )$ be the von Neumann entropy.
Then, we have
\begin{align}
|S_{AB}(U \rho U^\dagger)-S_{AB}(\rho)| \le (2 \log \chi) |C|
\end{align}
where $|C|$ is the number of qudits in $C$.
\begin{proof}
Let $\rho' = U \rho U^\dagger$.
\begin{align*}
& |S_{AB}(\rho') - S_{AB}(\rho)|\\
&=|S_{AB}(\rho') - S_{ABC}(\rho') + S_{ABC}(\rho') - S_{AB}(\rho)|\\
&=|S_{AB}(\rho') - S_{ABC}(\rho') + S_{ABC}(\rho) - S_{AB}(\rho)|\\
&\le |S_{AB}(\rho') - S_{ABC}(\rho')| + |S_{ABC}(\rho) - S_{AB}(\rho)|\\
&\le S_{C}(\rho') + S_C(\rho) \le (2 \log \chi) |C|
\end{align*}
In the second inequality, we used the subadditivity of entropy.
\end{proof}

The inequality is saturated by the swap operator.
If $A,B,C,D$ are single qubits, respectively, and $\psi$ consists of two pairs of singlets
in $AB$ and $CD$, then $S_{AB}(\psi) = 0$.
Swapping $B$ and $C$, we have $S_{AB}(\psi') = 2 \log 2$.
The lemma implies that a finite depth quantum circuit can only generate entanglement
between two regions along the boundary.

We wish to consider the entanglement entropy $S_0( \ket \psi ) = S(\rho)$, where $\rho = \tr_{B^c} (\ket \psi \bra \psi)$,
between a (hyper)cubic region $B$ of linear size $L$ and its complement
of a branching MERA state $\ket \psi$.

By definition, $\ket \psi$ accompanies entanglement RG transformations $U_\tau$ ($\tau = 1,2,\ldots$).
$U_1 \ket \psi$ is either a tensor product of one or more states $\ket{\psi_1^1},\ket{ \psi_1^2}, \ldots, \ket{\psi_1^b}$ ($b \ge 1$)
each of which is living on a coarser lattice (branch),
or some entangled state of those.
To be concrete, suppose the density of degrees of freedom decreases
by a factor of $2^D$ on the coarser lattice.
The number $b$ of branches should be $\le 2^D$.

Let $\rho_1^{(1)}, \ldots, \rho_1^{(b)}$ be reduced density matrices of $U_1 \ket \psi$ for the corresponding region $B_1^{i}$ on each branch.
Each $B_1^i$ contains $(L/2)^D$ qudits. By the lemma and the subadditivity of entropy,
we have
\begin{align}
 S(\rho) 
 & \le S( \tr_{B^c} U_1 \ket \psi \bra \psi U_1^\dagger ) + c |\partial B| \nonumber \\
 & \le S( \rho_1^{(1)} ) + \cdots + S(\rho_1^{(b)}) + c |\partial B|   \label{eq:recursion}
\end{align}
where $c$ is a constant depending only on the detail of the circuit $U_1$'s locality property.
Here, $|\partial B|$ is the number of qudits outside $B$ but within the range of $U_1$ from $B$.
So, $c | \partial B | \le (2\log \chi) 2D (L+2)^{D-1}$ if $U_1$ is of depth 1 and range 2.
One can iterate the inequality Eq.~\eqref{eq:recursion} with $B_1^{i}$ in place of $B$.
\begin{equation}
 S(\rho) \le \sum_{i=1}^{b_N} S(\rho_N^{(i)}) +  c' \sum_{n=0}^{N-1}  b_n \left( \frac{L}{2^n}\right)^{D-1}
 \label{eq:entropy-bound}
\end{equation}
for any $N \ge 0$ where $b_n$ is the total number of all branches,
and $\rho_N^{(i)}$ is the reduced density matrix of $U_N U_{N-1} \cdots U_1 \ket \psi$
for the region $B_N^{(i)}$ of linear size $L/2^N$ on branch $i$.
In particular, $b_0 = 1$ and $b_1 = b$ above.
In a usual MERA, we have $b_n = 1$ for all $n$.
The constant $c'$ only depends on $\chi$ and the details of the depth and range of circuits $U_1, \ldots, U_N$.

An appropriate $N$ must be chosen in order for Eq.~\eqref{eq:entropy-bound} to be useful.
A straightforward choice is such that $B_N^{(i)}$ contains a constant number of qudits,
i.e., $N = \lfloor \log_2 L \rfloor$.
Then, $\rho_N^{(i)}$ is a density matrix of a constant number of qudits, so $S(\rho_N^{(i)}) = O(\log \chi)$.
Eq.~\eqref{eq:entropy-bound} finally implies
\begin{align}
 S(\rho) 
& \le O(\log \chi) \sum_{n=0}^{\lfloor \log_2 L \rfloor} b_n  \left( \frac{L}{2^n}\right)^{D-1}.
\end{align}
Specializing, we get
\begin{align}
S(\rho) & = \begin{cases}
O(L^{D-1}) & \text{if $b_n = b^n < (2^{D-1})^n$},\\
O(L^{D-1} \log L) & \text{if $b_n = (2^{D-1})^n$},\\
O \left(L^{\log_2 (b/2^{D-1})} \right)& \text{if $b_n = b^n > (2^{D-1})^n$}.
\end{cases}
\end{align}
The number $2$ is of course the linear size of a superblock, and can be replaced by any positive integer.

\end{document}